\documentclass[cameraready]{Interspeech}

\title{SAM: A Mamba-2 State-Space Audio-Language Model}

\author[affiliation={1}, orcid=0009-0009-3576-2770]{Taehan}{Lee}
\author[affiliation={1}, orcid=0009-0001-6348-5397]{Jaehan}{Jung}
\author[affiliation={1}, orcid=0000-0003-2981-0800, correspondingauthor]{Hyukjun}{Lee}

\address{
    $^1$ Sogang University, South Korea
}

\email{alpaca@sogang.ac.kr, jhjung22@sogang.ac.kr, hyukjunl@sogang.ac.kr}


\usepackage{comment}
\usepackage{multirow}
\usepackage{booktabs}
\usepackage{setspace}
\usepackage[table]{xcolor}
\usepackage{colortbl}
\usepackage{hyperref}
\definecolor{lightgray}{gray}{0.92}
\definecolor{taucolor}{RGB}{0,100,230}
\definecolor{coscolor}{RGB}{230,100,0}
\definecolor{taubg}{RGB}{235,245,255}  
\definecolor{cosbg}{RGB}{255,242,230}  

\usepackage[normalem]{ulem}

\begin{document}

\maketitle

\begin{abstract}
We present SAM, a State-space Audio-language Model that integrates an audio encoder with a Mamba-2 backbone.
SAM-2.7B achieves 21.1 mAP on AudioSet and 17.6 SPICE on AudioCaps, matching or surpassing larger 7B transformer-based models with fewer parameters.
We further provide the first systematic, representation-level analysis of how SSMs interact with audio encoder outputs:
(1) joint audio encoder finetuning is essential, supported by accuracy gains and observed adaptation of token representation rank and similarity across different SSM sizes;
(2) despite linear scaling, SSMs benefit more from compact, information-rich audio token representations than from excessively long token sequences; and
(3) incorporating instruction-following supervision substantially improves reasoning ability, boosting MMAU-Sound accuracy from 22.8 to 56.8.
Through comprehensive experiments and analysis, we establish practical design principles for SSMs as strong, scalable backbones for audio-language models.
\end{abstract}

\section{Introduction}
\label{sec:introduction}
In recent years, Audio Language Models (ALMs) that combine Transformer-based~\cite{transformer} language models with audio encoders have achieved notable results on diverse audio understanding tasks~\cite{ltu, gama, af2, af3, qwen3, mellow}.
However, their core network module, the transformer, has computational requirements that scale quadratically with sequence length due to the attention mechanism.
In language modeling, State Space Models (SSMs) such as Mamba~\cite{mamba1, mamba2} have emerged as efficient alternatives to transformers.
Recently, several works have demonstrated Mamba's capability as an LLM backbone for image understanding tasks~\cite{mlmamba, vlmamba, shakingupvlm}.
In this work, we present a \textbf{S}tate-space \textbf{A}udio-language \textbf{M}odel (\textbf{SAM}) and show that it can match or surpass larger transformer-based ALMs with fewer parameters.
\textcolor{black}{The most closely related work} is ssLALM~\cite{mambaaac}, which replaces the transformer LLM with Mamba-1 following LTU's training pipeline.
Unlike previous work, beyond switching the backbone LLM to a newer SSM (Mamba-2), we provide our \textcolor{black}{key findings}:
\begin{itemize}
    \item \textcolor{black}{\textbf{SAM-2.7B achieves 21.1 mAP on AudioSet and 17.6 SPICE on AudioCaps}, matching or surpassing larger 7B transformer-based ALMs with fewer parameters}. 
    \textcolor{black}{This trend holds across multiple SSM scales and training hyperparameters, validating Mamba-2 as a strong ALM backbone.}
    \item \textcolor{black}{\textbf{Joint audio-encoder finetuning is essential for SSMs.} Beyond accuracy gains, we observe that the audio encoder produces lower rank and more similar token representations for smaller SSMs, suggesting adaptation to the SSM’s reduced capacity to integrate audio information.}
    \item \textcolor{black}{\textbf{SSMs benefit more from rich yet compact audio token representations} than from simply exploiting their linear-time and memory scaling with respect to sequence length.}
    \item \textcolor{black}{\textbf{Structured Binary Question (BQ) and Multiple-Choice Question (MCQ) supervision unlocks SSMs' audio reasoning capability}, boosting MMAU-Sound accuracy from 22.8 to 56.8 (+34 points) and outperforming the transformer-based Gemma3n-4B baseline.} 
\end{itemize}
The code is publicly available at \url{https://github.com/sam-audio-language-model/sam}

\begin{figure*}[ht]
    \vspace{-10pt}
    \centering
    \includegraphics[width=0.9\linewidth]{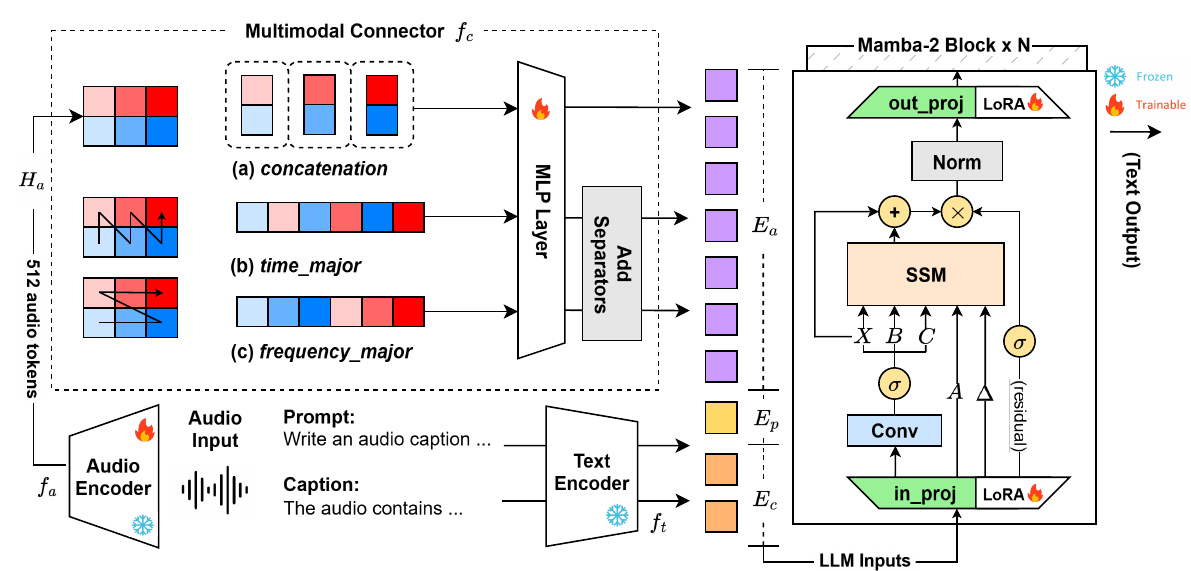}
    \caption{Overall architecture of our SSM-based Audio-language Model (SAM).}
    \label{fig:model}
\end{figure*}

\section{Model Architecture}

\label{sec:sam}
As shown in Figure~\ref{fig:model}, our model follows a standard multimodal LLM architecture, comprising an audio encoder $f_a$, a text encoder $f_t$, a connector $f_c$, and a Mamba-2 LLM.
Given a mel-spectrogram $x$, the audio encoder produces audio tokens $H_a=f_a(x)$, which are projected by the connector into audio embeddings $E_a=f_c(H_a)$.
The text encoder maps the prompt $p$ and caption $c$ to embeddings $E_p=f_t(p)$ and $E_c=f_t(c)$, respectively.
During training, the LLM uses the concatenated sequence $[E_a, E_p, E_c]$, while at inference it uses only $[E_a, E_p]$.
We train our models with an auto-regressive next-token cross-entropy loss over the ground-truth caption embeddings $E_c$.
\begin{gather}
    h_t = \bar A h_{t-1} + \bar B x_t, \quad y_t = C^\top h_t \tag{1} \label{eq:1}\\
    \overline{A} = \exp(\Delta A), \quad 
    \overline{B} = (\Delta A)^{-1}\bigl(\exp(\Delta A) - I\bigr)\cdot \Delta B \tag{2} \label{eq:2} \\
    \overline{K} = \bigl(C\overline{B}, C\overline{A}\overline{B}, \ldots, C\overline{A}^{N}\overline{B}\bigr),\quad 
    y = x * \overline{K}, \tag{3} \label{eq:3}
\end{gather}

\subsection{SSM}

Structured State Space Models (SSMs) in discrete time transform a 1D input sequence $ x\in \mathbb{R}^T \mapsto y \in \mathbb{R}^T$ as in (\ref{eq:1}).
The discretized parameters $(\bar A,\bar B)$ are obtained from the continuous-time parameters $(A, B)$ and a step size $\Delta \in \mathbb{R}$ via the zero-order hold (ZOH) discretization rule in (\ref{eq:2}).
An SSM can be computed either in a recurrent form (\ref{eq:1}) or in parallel via a convolution mode (\ref{eq:3}).
S4D~\cite{s4d} constrains $A$ to be diagonal, enabling efficient computation.
SSM-based LLMs map a multi-dimensional input sequence $x\in \mathbb{R}^{T\times D} \mapsto y \in \mathbb{R}^{T \times D}$. Mamba~\cite{mamba1} improves language-modeling performance with input-dependent SSM parameters for $x_t\in\mathbb{R}^D$: $\Delta_t^{(d)}=\text{softplus}(W_\Delta^{(d)} x_t+\beta_t)$, $B_t=W_Bx_t$, $C_t=W_c x_t$.
Mamba also introduces hardware-efficient computation kernels based on parallel selective scan.
A Mamba block consists of a per-channel SSM, a 1D convolution, and a gated MLP.

Mamba-2~\cite{mamba2} groups channels into heads that share a single SSM and reformulates the SSM in matrix-multiplication form with block decomposition, reporting $2\text{-}8\times$ faster training while matching transformer models in performance.
In contrast to Mamba’s per-channel diagonal matrix \(A\), Mamba-2 uses a scalar-times identity \(A_t=a_t I\) per head and increases the state size \(N\) to improve language modeling capability.
For additional details, please refer to the original paper~\cite{mamba2}.
We use Mamba-2 (130M, 780M, and 2.7B) models pretrained on the Pile~\cite{pile} dataset as our LLM backbone.

\subsection{Audio encoder}
We use EAT-base~\cite{eat} (88M parameters) finetuned on AudioSet~\cite{audioset} as our audio encoder for three reasons:
(1) it achieves strong performance on audio classification tasks and has demonstrated effectiveness in prior ALMs~\cite{slamaac, dcaseaac};
(2) its AudioSet accuracy (48.6 mAP) closely matches that of DASS (48.9 mAP), the audio encoder used in our ssLALM baseline with a Mamba-1 backbone; and
(3) its output shape enables an ablation study on whether providing the SSM with longer, uncompressed audio tokens is beneficial.
EAT uses a CNN patch-embedding layer followed by 12 ViT blocks, producing 512 audio tokens with a feature dimension of 768.

\subsection{Multimodal connector}
We adopt a two-layer MLP as the multimodal connector, as it has shown competitive performance against more complex designs~\cite{mellow}.
We compare three connector designs for producing the audio embeddings $E_a$.
(a) \textit{concatenation}: reshape the 512 initial tokens into a $(64, 8)$ time-frequency representation and concatenate them along the frequency axis, yielding 64 tokens with a 6144-dimensional embedding.
These tokens are projected by the MLP, resulting in a final length of $|E_a|=64$.
Unlike transformers, which enable global interactions among tokens via attention, an SSM updates its hidden state recurrently; thus, token ordering directly affects how audio information is processed.
Motivated by this property, we propose two novel connector designs:
(b) \textit{time\_major}: rearrange tokens along the time axis to preserve temporal continuity in the SSM state updates; and
(c) \textit{frequency\_major}: rearrange tokens along the frequency axis to retain spectral locality.
Inspired by prior works addressing Mamba's limited positional awareness~\cite{shakingupvlm, vmamba}, we insert separator token embeddings (``\&\&'') to mark boundaries between time steps or frequency bands, enabling the SSM to better preserve structural cues in the embedding sequence.
The resulting sequence lengths are $|E_a|=64\times(8+1)=576$ for (b) and $|E_a|=(64+1)\times8=520$ for (c).

\begin{table*}[ht]
\setlength{\tabcolsep}{8pt}
\caption{Quantitative comparison between baseline models and our models of different sizes on audio description tasks.
$r$: LoRA rank; (a): \textit{concatenation}; (b): \textit{time\_major}; (c): \textit{frequency\_major}. 
Each experiment is denoted as \texttt{E\#}.
\textbf{Bold} and \underline{underline} indicate the best and second-best results, respectively.}
\label{tab:result}
\centering
\resizebox{\textwidth}{!}{
\renewcommand{\arraystretch}{0.848}
\begin{tabular}{@{}c l c c c c c c c c c c c c}
\toprule
\multicolumn{2}{c}{\multirow{2}{*}{\textbf{Model}}} & \multicolumn{2}{c}{\textbf{Config}} & 
\textbf{ESC50} &
\textbf{DCASE} &
\textbf{VS} &
\textbf{TUT} &
\textbf{BJO} & 
\textbf{VGG} &
\textbf{FSD} &
\textbf{AudioSet} &
\textbf{AudioCaps} &
\textbf{Clotho}\\
\cmidrule(lr){3-14}
& & \textbf{$r$} & \textit{Conn.} &
\multicolumn{1}{c}{(Acc)} &
\multicolumn{1}{c}{(Mi-F1)} &
\multicolumn{1}{c}{(Acc)} &
\multicolumn{1}{c}{(Acc)} &
\multicolumn{1}{c}{(Acc)} &
\multicolumn{1}{c}{(Acc)} &
\multicolumn{1}{c}{(mAP)} &
\multicolumn{1}{c}{(mAP)} &
\multicolumn{1}{c}{(SPICE)} &
\multicolumn{1}{c}{(SPICE)} \\
\midrule

\multicolumn{14}{l}{\textbf{\textit{Baseline Audio Language Models}}} \\
\multicolumn{2}{l}{LTU-7B~\cite{ltu}} & 8 & Linear & 83.1 & 45.9 & 55.6 & 32.5 & \textbf{69.9} & 50.3 & 46.3 & 18.7 & 17.0 & 11.9 \\
\multicolumn{2}{l}{GAMA-7B~\cite{gama}} & 8 & QFormer & 82.6 & 38.4 & 52.4 & 21.5 & 69.5 & \textbf{52.2} & \textbf{47.8} & 19.2 & \textbf{18.5} & \textbf{13.5} \\
\multicolumn{2}{l}{ssLALM-2.8B~\cite{mambaaac}} & 8 & Linear & \textbf{86.8} & \textbf{47.9} & \textbf{61.2} & \textbf{35.9} & 61.0 & 51.0 & 47.7 & \textbf{19.4} & 17.7 & 11.7 \\

\midrule
\texttt{E} & \multicolumn{13}{l}{\textbf{\textit{Our SSM-based Audio-language Models (SAMs)}}} \\
\texttt{1} & SAM-130M & 8 & (a) & 79.7 & 42.8 & 55.7 & 29.2 & 49.2 & 51.8 & 47.0 & 19.7 & 14.7 & 10.5 \\
\texttt{2} & SAM-780M & 8 & (a) & 83.9 & 45.5 & 56.3 & 29.4 & 57.6 & 54.5 & 47.0 & 20.3 & 16.2 & 11.5 \\
\texttt{3} & SAM-2.7B & 8 & (a) & 87.1 & \uline{48.7} & 63.2 & 32.8 & \textbf{69.1} & 56.6 & 48.4 & 20.8 & 17.3 & 12.4 \\
\texttt{4} & SAM-130M & 256 & (a) & 83.3 & 47.3 & 59.8 & \uline{33.3} & 62.3 & 53.5 & 46.7 & 19.9 & 16.6 & 12.0 \\
\texttt{5} & SAM-780M & 256 & (a) & 87.7 & 47.7 & 61.9 & 31.7 & 49.2 & 56.1 & 48.8 & \textbf{21.1} & 16.9 & 12.2 \\
\rowcolor{lightgray}
\texttt{6} & SAM-2.7B & 256 & (a) & \uline{89.7} & \uline{48.7} & 70.9 & 31.6 & \uline{65.3} & \textbf{57.0} & \textbf{49.2} & \textbf{21.1} & \uline{17.6} & 11.8 \\

\midrule
\multicolumn{14}{l}{\textbf{\textit{Ablation 4.1 - Audio encoder excluded from training}}} \\
\texttt{7} & SAM-130M & 256 & (a) & 76.5 & 43.9 & 48.2 & 31.6 & 56.8 & 48.1 & 39.9 & 18.2 & 16.7 & 11.1 \\
\texttt{8} & SAM-780M & 256 & (a) & 84.3 & 46.5 & 62.1 & 32.5 & 52.5 & 53.3 & 45.0 & 20.4 & 16.1 & 11.8 \\
\texttt{9} & SAM-2.7B & 256 & (a) & 86.4 & \textbf{48.9} & 67.3 & 31.8 & 58.9 & 54.3 & 44.5 & \uline{21.0} & 16.9 & 11.8 \\

\midrule
\multicolumn{14}{l}{\textbf{\textit{Ablation 4.2 - Providing uncompressed audio tokens to SSM}}} \\
\texttt{10} & SAM-130M & 256 & (b) & 83.2 & 47.3 & 64.1 & 30.9 & 46.6 & 53.1 & 46.3 & 18.8 & 16.8 & 12.1 \\
\texttt{11} & SAM-130M & 256 & (c) & 84.5 & 47.3 & 63.6 & 32.0 & 62.7 & 52.9 & 46.3 & 19.1 & 16.4 & 12.2 \\
\texttt{12} & SAM-780M & 256 & (b) & 88.3 & 46.6 & 68.0 & \textbf{33.2} & 61.0 & 55.7 & 48.6 & 20.0 & 17.2 & 12.3 \\
\texttt{13} & SAM-780M & 256 & (c) & 87.6 & 47.4 & 64.1 & 32.8 & 58.1 & 56.2 & 48.4 & 20.5 & 17.2 & 12.3 \\
\texttt{14} & SAM-2.7B & 256 & (b) & \textbf{90.0} & 48.3 & \uline{71.0} & 32.4 & 64.4 & \uline{56.7} & 48.9 & \textbf{21.1} & \textbf{17.8} & \textbf{13.0} \\
\texttt{15} & SAM-2.7B & 256 & (c) & 88.2 & 48.1 & \textbf{72.2} & 32.4 & 64.4 & \uline{56.7} & \uline{49.0} & 20.7 & 17.2 & \uline{12.8} \\

\midrule
\multicolumn{14}{l}{\textbf{\textit{Ablation 4.3 - Enhancing reasoning ability with more instruction-following dataset (OpenReasonAQA)}}} \\
\texttt{16} & SAM-130M & 256 & (a) & 82.7 & 44.4 & 65.8 & 31.5 & 57.6 &	51.7 & 44.0 & 19.0 & 17.2 &	12.2 \\
\texttt{17} & SAM-780M & 256 & (a) & 86.2 &	46.7 & 62.5 & 30.7 & 48.3 &	54.4 & 46.6 & 20.7 & 17.5 &	11.8 \\
\texttt{18} & SAM-2.7B & 256 & (a) & 87.0 &	46.7 & 70.9 & 27.7 & 57.2 &	55.6 & 46.7 & 20.8 & 17.4 &	11.3 \\

\bottomrule
\end{tabular}
}
\end{table*}

\section{Experiment}
\subsection{Model training}
We train our models on the OpenAQA dataset, which contains 1.9M closed-ended and 3.7M open-ended QA pairs, following the 4-stage LTU curriculum learning strategy.
We apply LoRA~\cite{lora} adapters to the \textit{in\_proj} and \textit{out\_proj} layers of each Mamba-2 block with $\alpha=2r$, as it has been shown effective for parameter-efficient adaptation of SSM-based models~\cite{mamba-peft-iclr25, mamba-peft-icml25}.
To accelerate training, we employ FlashAttention-2~\cite{flashattention-2} for the self-attention layers in the EAT encoder. 
Depending on model size, training takes approximately 0.5 to 2 days using mixed precision (bfloat16) on two NVIDIA RTX 4090 GPUs.

\subsection{Evaluation method}
We follow LTU evaluation protocol: (i) zero-shot accuracy on classification tasks (ESC-50~\cite{esc50}, DCASE2017 Task 4~\cite{dcase2017}, VocalSound~\cite{vocalsound}, TUT-2017~\cite{tut2017}, Beijing Opera~\cite{bjo}, VGGSound~\cite{vgg}, and FSD-50K~\cite{fsd50k}) and (ii) SPICE~\cite{spice} on captioning tasks (AudioCaps~\cite{audiocaps} and Clotho~\cite{clotho}) with greedy decoding.
To evaluate (i), we encode both generated captions and ground-truth labels with the CLAP~\cite{clap} text encoder and compute their cosine similarity in the joint audio-text embeddings.

\subsection{Results}
Table~\ref{tab:result} presents a comprehensive comparison between our models and prior ALMs trained on OpenAQA.
Our flagship model, SAM-2.7B (\texttt{E6}), outperforms larger transformer-based ALMs and the prior state-space model on multiple benchmarks.
Furthermore, our smaller models (\texttt{E4-5}) demonstrate competitive performance against substantially larger baseline ALMs on audio description tasks.
Increasing the LoRA rank from $r=8$ to $256$ (\texttt{E1-3} vs. \texttt{E4-6}) consistently improves performance, with larger gains for smaller SSMs.
Notably, even with a larger LoRA rank ($r=256$), Mamba-2 requires approximately 20\% less training time than Mamba-1 with $r=8$, owing to Mamba-2's matrix-multiplication-based computation kernel.
We also provide a qualitative comparison of audio descriptions from our models and prior ALMs in Table~\ref{tab:qualitative}.

\begin{table}[ht]
\renewcommand{\arraystretch}{0.9}
    \centering
    \small
    \caption{Qualitative comparison between baseline models and our models of different SSM sizes.}

    \begin{tabular}{p{0.95\columnwidth}}
    \toprule

    \textbf{Reference: A construction vehicle engine running and water splashing as wood crackles and snaps followed by electronic beeping and a man talking.} \\
    
    \midrule 
    \textbf{LTU-7B}: A boat is being rowed with splashing water and creaking noises. \\
    \textbf{GAMA-7B}: A man speaks amidst the sounds of a bus and a sliding door, possibly giving instructions or commentary on the journey. \\ \midrule
    \textbf{SAM-130M}: A beep and a man speaking. \\
    \textbf{SAM-780M}: A truck engine is running and a man speaks. \\
    \textbf{SAM-2.7B}: A large motor vehicle engine is running, metal clanking occurs, and an adult male speaks. \\
    \bottomrule
    \end{tabular}
    \label{tab:qualitative}
\end{table}

\section{Ablation Studies}
\label{sec:ablation}

\subsection{Why do SSMs benefit from finetuning audio encoders?}
\begin{figure}[ht]
    \centering
    \includegraphics[width=\columnwidth]{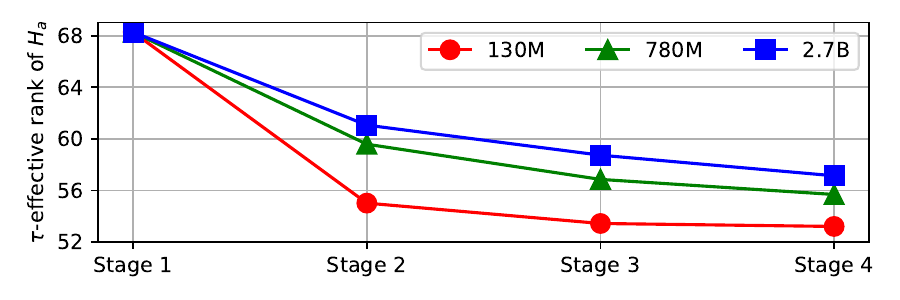}
    \caption{$\tau$-effective rank across training stages by model size.}
    \label{fig:erank_graph}
\end{figure}

Several ALMs finetune the audio encoder end-to-end~\cite{ltu, gama, af3} to improve task performance.
In contrast, other approaches freeze the audio (or vision) encoder~\cite{mellow, mlmamba, shakingupvlm}, motivated not only by reducing computational cost but also by mitigating catastrophic forgetting~\cite{blip2} or preserving modality-specific representations~\cite{facegpt}.
We compare both strategies on the audio description tasks and observe that jointly training the audio encoder consistently improves performance (\texttt{E4-6} vs. \texttt{E7-9}).
This observation suggests that the benefit of end-to-end training may arise from improved alignment between the encoder representations and the SSM’s sequential token processing mechanism.
\textcolor{black}{Unlike transformers, which can attend to all previous token embeddings via self-attention at each step}, \textcolor{black}{SSM-based decoders must summarize the entire prefix through a fixed-dimensional recurrent state.
This induces \textcolor{black}{a state-capacity bottleneck}: each incoming audio token must be integrated online into the state, and information that is not preserved at that moment cannot be recovered later by re-attending to earlier tokens.
Accordingly, the amount of information that can be preserved and integrated through this recurrent state depends on model scale and is more constrained for smaller SSMs.
To analyze the audio encoder's adaptation under this bottleneck, we measure the $\tau$-effective rank of extracted audio tokens as a proxy for embedding space utilization~\cite{tauedim}:}
\begin{gather}
    rank_\tau(H_a)=\text{argmin}_k \left( \frac{\sum_{i=1}^k{\sigma_i^2}}{\sum_{i=1}^{d}{\sigma_i^2}}\geq \tau \right) \tag{4} \label{eq:4}
\end{gather}
where $\sigma_i$ is the $i$-th largest singular value of $H_a$.
We set $\tau=0.95$ and compute $rank_\tau(H_a)$.
Table~\ref{tab:tauerank} shows a clear size-dependent shift in audio token characteristics under end-to-end training: smaller SSMs are paired with more reduced token representations (higher token-to-token cosine similarity and lower $\tau$-effective rank).
Figure~\ref{fig:erank_graph} shows how $rank_\tau(H_a)$ changes across training stages and SSM scales.

\begin{table}[ht]
\centering
\setlength{\tabcolsep}{3.5pt}
\renewcommand{\arraystretch}{1.1}
\caption{Analysis of audio tokens by different model sizes.}

\begin{tabular}{c | c c c }
\hline
\multirow{2}{*}{\textbf{Size}} 
& \textbf{SAM-(a)} & \textbf{SAM-(b)} & \textbf{SAM-(c)} \\
\cline{2-4}
& \multicolumn{3}{c}{$\tau$-effective rank / cosine-similarity} \\
\hline
130M & 53.2 / 0.47 & 39.5 / 0.53 & 40.3 / 0.54 \\ 
780M & 55.7 / 0.40 & 50.0 / 0.47 & 49.7 / 0.47 \\
2.7B  & 57.1 / 0.35 & 51.0 / 0.45 & 51.2 / 0.46 \\
\bottomrule
\end{tabular}
\label{tab:tauerank}
\end{table}

To further analyze this phenomenon, we swap in an audio encoder finetuned at a different SSM size (\texttt{E4-6}), freeze it, and retrain SAM-130M from scratch.
Specifically, we train SAM-130M using a frozen audio encoder initialized from the converged checkpoints of \texttt{E4-6}.
\textcolor{black}{If encoder finetuning only improved generic acoustic representations, we would expect encoders trained with larger SSMs to produce better representations.
Instead, the encoder-swap results in Table~\ref{tab:audioswap} show that size-matched encoder achieves the strongest performance in most cases.
This supports our interpretation: size-mismatched encoders degrade SAM-130M, indicating that the learned audio token representations are not universally optimal but are better aligned with the SSM’s effective integration capacity.}

\begin{table}[ht]
\centering
\caption{Performance of SAM-130M trained with frozen audio encoders transferred from E4--E6.}
\begin{tabular}{r|cccc|cc}
\toprule
\multicolumn{1}{c|}{\textbf{Source\hspace{0.4em}$f_a$}} & \textbf{ESC} & \textbf{VGG} & \textbf{FSD} & \textbf{AS} & \textbf{AC} & \textbf{CL} \\
\hline
130M (\texttt{E4}) & 81.5 & 53.1 & 46.2 & 19.8 & 16.9 & 11.4 \\
780M (\texttt{E5}) & 80.2 & 52.8 & 45.6 & 19.8 & 16.9 & 11.5 \\
2.7B (\texttt{E6}) & 80.2 & 52.7 & 45.4 & 19.4 & 16.4 & 11.1 \\
\bottomrule
\end{tabular}
\label{tab:audioswap}
\end{table}

\subsection{Do SSMs benefit from uncompressed audio tokens?}

\begin{figure}[ht]
    \centering
    \includegraphics[width=\columnwidth]{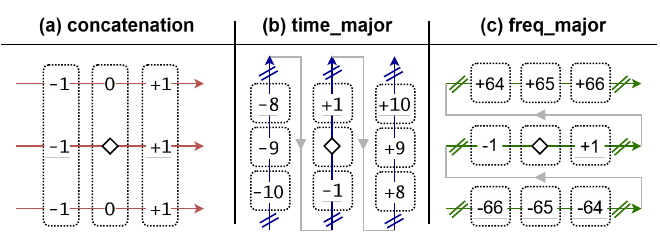}
    \caption{State update distance between adjacent audio tokens.}
    \label{fig:ablation}
\end{figure}

SSMs' time and memory costs scale linearly with sequence length.
This property raises the question of whether SSMs can benefit from using long, uncompressed audio tokens.
To investigate this, we conduct a comparative analysis of different multimodal connector designs.
In setup (a), the concatenated 6,144-dimensional audio tokens are compressed via projection into the SSM's hidden dimension (768, 1,536, and 2,560 for the 130M, 780M, and 2.7B models, respectively).
In contrast, in setups (b) and (c), the full audio token sequence is fed into the SSM using a connector that either preserves the original embedding dimension or further expands it.
We observe that using uncompressed audio tokens (b-c: \texttt{E10-15}) does not consistently outperform compressed tokens (a: \texttt{E4-6}).
Although the embedding dimension is not reduced in (b) and (c), longer token sequences impose a greater burden on the SSM.
As illustrated in Figure~\ref{fig:ablation}, for each audio token, information from neighboring tokens needs to be maintained across more sequential state updates than in (a).
Table~\ref{tab:tauerank} further shows that models with uncompressed audio tokens (b, c) exhibit lower $\tau$-effective rank than compressed models (a), particularly in smaller models, indicating less effective utilization of representational capacity.
These findings reveal that SSMs benefit more from compact and information-rich token representations than from relying on excessively long audio token sequences, despite SSMs' linear-time/memory scalability.
\subsection{How to enhance reasoning ability of Mamba-2 SSMs?}
\begin{table}[ht]
\renewcommand{\arraystretch}{0.9}
\centering
\setlength{\tabcolsep}{3pt}
\caption{MMAU-v05.15.25 benchmark results.}
\begin{tabular}{c | r r | r r | r r}
\toprule
\multirow{2}{*}{\textbf{Model}} 
& \multicolumn{2}{c|}{\textbf{Sound}} 
& \multicolumn{2}{c|}{\textbf{Music}} 
& \multicolumn{2}{c}{\textbf{Speech}} \\
\cline{2-7}
& \textbf{mini} & \textbf{base} & \textbf{mini} & \textbf{base} & \textbf{mini} & \textbf{base} \\
\hline
Gemma 3n-4B  & 55.86	& 50.27	& 56.89	& 53.20	& 61.26	& 62.13 \\
Gemma 3n-2B  & 51.35	& 47.47	& 52.10	& 51.63	& 52.22	& 57.07	\\
GAMA-7B     & 31.83	& 30.73	& 17.71	& 17.33	& 12.91 & 16.97 \\
LTU-7B      & 22.52	& 25.86	& 9.69  & 12.83	& 17.71	& 16.37	\\
\hline
SAM-2.7B    & 24.92 & 22.83 & 16.47 & 15.87 & 3.30 & 4.70 \\
SAM-780M    & 9.01  & 7.80  & 7.49 & 6.13 & 2.70 & 2.10 \\
SAM-130M    & 11.41 & 11.67 & 5.99 & 7.67 & 5.11 & 4.77 \\
\hline
\rowcolor{lightgray}
SAM+OR-2.7B  & 61.86 & 56.77 & 48.80 & 50.43 & 31.23 & 27.43 \\
SAM+OR-780M  & 58.86 & 59.73 & 43.71 & 46.63 & 30.33 & 33.00 \\
SAM+OR-130M  & 48.95 & 46.93 & 40.12 & 41.47 & 20.72 & 23.93 \\
\bottomrule
\end{tabular}
\label{tab:mmau}
\end{table}
Beyond audio description tasks, recent ALM studies have increasingly evaluated instruction comprehension and audio reasoning through binary questions (BQ) and multiple-choice questions (MCQ).
To explicitly investigate whether stronger instruction-following supervision improves audio reasoning, we construct \textbf{O}pen\textbf{R}easonAQA, based on ReasonAQA~\cite{mellow}, which provides detailed QAs and MCQs derived from AudioCaps and Clotho.
Specifically, we replace caption-style open-ended questions with the structured BQ/MCQ reasoning questions from ReasonAQA. To maintain a comparable training scale, we repeat them 5 times, resulting in 3.8M pairs.
As shown in Table~\ref{tab:mmau}, Mamba-2 trained with intensive reasoning supervision (SAM+OR) consistently improves performance over SAM and SAM+OR-2.7B surpasses Gemma3n-4B~\cite{gemma3n} on MMAU~\cite{mmau} Sound benchmarks.
These results suggest that structured instruction-following supervision is particularly effective in strengthening audio reasoning ability 
\textcolor{black}{across all model scales, highlighting the importance of training data composition for reasoning tasks.}

\section{Conclusion}
In this work, we demonstrate that Mamba-2 exhibits strong audio-language modeling performance, achieving competitive results against larger transformer-based ALMs while using fewer parameters.
Through extensive experiments and analysis of SSM-specific properties, we provide practical insights for training and designing SSM-based audio-language models.
As future work, we plan to explore hybrid SSM–transformer architectures to further improve audio reasoning.

\section{Generative AI Use Disclosure}
We used a generative AI tool (ChatGPT, OpenAI) solely to assist with English proofreading and language polishing (e.g., grammar). The tool was not used to generate scientific contents or conclusions. All authors reviewed and edited the AI-suggested changes, and take full responsibility for the content and its submission.

\bibliographystyle{IEEEtran}
\bibliography{mybib}

\end{document}